\documentclass[prb,twocolumn,aps,floatfix,preprintnumbers,amsmath,amssymb]{revtex4}
\usepackage{natbib}
\usepackage{dcolumn}
\usepackage{graphicx}
\usepackage{subfigure}
\usepackage{hyperref}
\usepackage{verbatim}

\newcommand{\mnras}{Mon.~Not.~R.~Astron.~Soc.}

\newcommand{\phrd}{Phys.~Rev.~D.}
\newcommand{\jcap}{J.~Cosmol.~Astropart.~Phys.}

\newcommand{\be}{\begin{equation}}
\newcommand{\ee}{\end{equation}}
\newcommand{\bea}{\begin{eqnarray}}
\newcommand{\eea}{\end{eqnarray}}
\begin{document}
\title{Modern cosmology: Interactive computer simulations that use recent observational surveys}
\author{Jacob Moldenhauer$^1$\footnote{Electronic address: JMoldenhauer@fmarion.edu}, Larry Engelhardt$^1$, Keenan M. Stone$^1$, Ezekiel Shuler$^1$}
\affiliation{
$^1$Department of Physics and Astronomy, Francis Marion University, Florence, South Carolina 29506}
\date{\today}
%
%
\begin{abstract}
We present a collection of new, open-source computational tools for numerically modeling recent large-scale observational data sets using modern cosmology theory.  Specifically, these tools will allow both students and researchers to constrain the parameter values in competitive cosmological models, thereby discovering both the accelerated expansion of the universe and its composition (e.g., dark matter and dark energy).  These programs have several features to help the non-cosmologist build an understanding of cosmological models and their relation to observational data: a built-in collection of several real observational data sets; sliders to vary the values of the parameters that define different cosmological models; real-time plotting of simulated data; and $\chi^2$ calculations of the goodness of fit for each choice of parameters (theory) and observational data (experiment). The current list of built-in observations includes several recent supernovae Type Ia surveys, baryon acoustic oscillations, the cosmic microwave background radiation, gamma-ray bursts, and measurements of the Hubble parameter.  In this article, we discuss specific results for testing cosmological models using these observational data. These programs can be found at \url{http://www.compadre.org/osp/items/detail.cfm?ID=12406}.  

\end{abstract}
\pacs{Pacs here}
\maketitle

\section{Introduction}
In recent decades, the field of cosmology---both observational data and theoretical models---has provided two very significant insights regarding the nature of our universe. One discovery, which earned the 2011 Nobel Prize in Physics \cite{NobelPrize2011, Riess1998, Perlmutter1999}, is that our universe is not just expanding; it is expanding at a rate that is \emph{increasing} with time. (i.e., the matter in the universe is experiencing a repulsion that overcomes the attractive force of gravity.) The other major discovery is that---in order to successfully model the (numerous) recent astronomical observations---our universe must be composed of mostly dark matter and dark energy, with only 4\% ordinary matter (e.g., atoms) \cite{WMAP2010}.  Clearly, these results are of intrinsic interest and should be understood by people outside the field of cosmology. In fact, several recent articles in non-specialist journals have discussed these latest findings  \cite{PhysicsToday1, PhysicsToday2, SonegoTalamini}. Unfortunately, resources have not existed that allow the broader physics community (non-cosmologists) to appreciate how cosmological observations inform cosmological models, ultimately leading to these new insights.  In the present work, we seek to address this need.

In order to help non-cosmologists to understand how recent observations lead to the discoveries described above, we provide a medley of open-source, user-friendly cosmological modeling programs, which we will refer to as ``\texttt{COSMOEJS}''\cite{CosmoEJS}.  These programs allow the user to immediately become an amatuer cosmologist by fitting theoretical models to the actual experimental data and visually observing how well each model agrees with the various observational data. Key features of \texttt{COSMOEJS} include a built-in collection of several real observational data sets; sliders to vary the values of the parameters that define different cosmological models; real-time plotting of simulated data; and $\chi^2$ calculations of the goodness of fit for each choice of parameters (theory) and observational data (experiment). Taking advantage of these features, \texttt{COSMOEJS} has already been used with a variety of non-cosmologist audiences to bridge the gap between modern cosmology and mainstream physics.

In Sec.~\ref{sec:modeling} we describe some specific examples of modeling cosmology with experimental observations using \texttt{COSMOEJS}, and  Sec.~\ref{sec:summary} contains a summary of our results. We also provide a brief introduction to cosmology in the Appendix ~\ref{sec:background}, (a more detailed introduction is provided in supplementary material \cite{Observations}).  The relevant mathematical quantities and equations are introduced in subsection \ref{sec:mathematics} of the Appendix, and the relevant experimental observations are introduced in subsection \ref{sec:expobservations} of the Appendix.


\section{Modeling Cosmology}\label{sec:modeling}
\texttt{COSMOEJS} allows non-cosmologists to simulate the expansion of the universe using various models and to compare the simulated results to experimental observations.  Recent articles in non-specialized physics journals have discussed some of the complexities associated with modeling cosmology \cite{PhysicsToday1, PhysicsToday2, SonegoTalamini}, but these articles only considered specific scenarios from the small subset of cosmological models that permit exact analytical solutions. Moreover, it would be very difficult for a non-cosmologist to recreate or extend those results. Some web-based \cite{NedWright} and mobile device applications \cite{iphoneApp} do provide ``cosmology calculators" for calculating times and distances of simple models, but do not compare to data, or allow for the diversity of models contained \texttt{COSMOEJS}.  Very powerful numerical simulations for testing cosmological models and constraining the values of model parameters do already exist \cite{cosmomc, icosmo}, but these tools have a steep learning curve, making them impractical for use by non-specialists. \texttt{COSMOEJS} addresses all of these needs. It provides direct comparisons between theory and experiment in the form of both plots and numbers, and it accurately carries out the complex mathematics without requiring technical expertise from the user. This allows the user to focus on developing a high-level understanding of cosmology, without technical (mathematical and computational) distractions.

The process of using \texttt{COSMOEJS} is very straightforward, and we encourage the user to download the program from Ref.~\onlinecite{CosmoEJS} in order to recreate and modify the plots that are discussed later in this section.  Using \texttt{COSMOEJS} consists of five steps: (1) loading observational data, (2) selecting values for the model parameters, (3)  calculating and plotting theoretical observables, (4) assessing the goodness of fit both visually and numerically, and (5) plotting the expansion of the universe for the user-defined model. For step (1), the user can select up to 18 different experimental datasets, and a drop-down menu is provided to simplify the process of loading data (see subsection \ref{sec:expobservations} of the Appendix for a description of these experimental data) For step (2), users can use sliders to adjust parameter values, subsequently changing from one model to another. For step (3), the calculations are carried out using Romberg's method of approximating integrals, and the user can easily vary the number of partitions to test for convergence of the numerical integration. For step (4), each time that the user changes the model's parameters, several plots are generated; and for each plot, $\chi^2$ is automatically calculated to provide a quantitative measure of the goodness of fit. Finally, for step (5), the size of the universe can be plotted versus time to see what type of universe results from each set of parameters; i.e., is the expansion of the universe constant, or accelerating, or decelerating? 

\subsection{Fitting the model}\label{sec:fitting}
In this subsection, we demonstrate the modeling capabilities of \texttt{COSMOEJS} by comparing experimental data (astronomical observations) with theoretical curves for a few specific examples that do not permit analytical solutions.  The experimental data consist of measurements of Type 1a supernovae (SNe), the Hubble parameter, $H(z)$, gamma ray bursts (GRB), baryon acoustic oscillations (BAO), and the cosmic microwave background (CMB).  (Each of these observations is described in subsection \ref{sec:expobservations} of the Appendix.)  The simulated data in this section consist of three physically different models, each described by a different set of parameter values. Specifically, in each model, the universe is chosen to have the same expansion rate today, but different fractions of matter, $\Omega_m$, and dark energy, $\Omega_\Lambda$: $\{\Omega_m,\,\Omega_\Lambda\}=\{0.01,\,0.99\},\,\{0.27,\,0.73\}$, and $\{1.0,\,0.0\}$. 

Throughout this section data are plotted versus redshift.   
It is important to note that redshift, $z$, can be used as a measure of time.  \footnote{The redshift, $z$, is defined as $z \equiv (\lambda_{obs}-\lambda_{emit})/\lambda_{emit}$, where $\lambda_{obs}$ and $\lambda_{emit}$ are the observed and emitted wavelengths, respectively, for the source\cite{BergstromGoobar}.}  Light from nearby objects experiences very little redshift ($z \approx 0$), and this light was also emitted very recently (in terms of cosmological timescales).  Light from far-away objects experiences a larger redshift, and this light was emitted longer ago.  \footnote{This is only comparatively speaking, for a precise calculation of time from redshift, a model must be chosen, see Sec. ~\ref{sec:interpretation}.}  We take advantage of this redshift/time relationship in multiple ways.
Given a theoretical model [Eq. (\ref{eq:FriedmannEquationH}) in subsection \ref{sec:mathematics} of the Appendix], \texttt{COSMOEJS} uses redshift values to calculate and display both the age of the universe today and the ``look-back" time, which refers to how long ago the light was emitted that is observed to have a certain redshift.  Also, the experimental data are measured using redshift (x-axis), which provides a means of ``dating" these observational data once a particular model has been selected. 

\begin{figure}[h!]

\begin{center}
\includegraphics[scale=0.28]{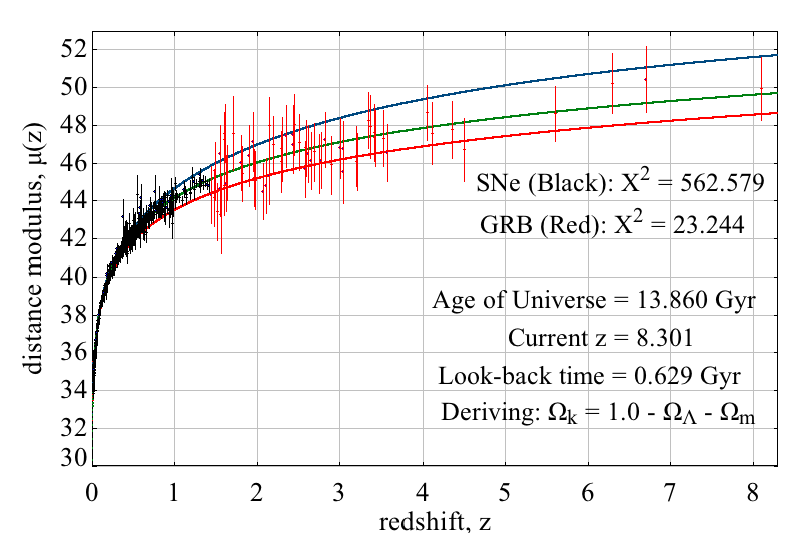}

\caption{\label{PlotsSNeGRB}  (color online)
Supernovae Type Ia and Gamma Ray Bursts versus redshift. \texttt{COSMOEJS} output showing three different models (curves) and two different experimental data sets, SNe (black) and GRB (red). The three models differ \emph{only} in their fractional matter and dark energy densities, $\{\Omega_m,\,\Omega_\Lambda\}=\{0.01,\,0.99\},\,\{0.27,\,0.73\},\,\{1.0,\,0.0\}$ correspond to top (blue), middle (green), and bottom (red), respectively.  (Note: Labels are draggable in \texttt{COSMOEJS}.)}

\end{center}
\end{figure}

In Fig.~\ref{PlotsSNeGRB}, distance modulus, $\mu$, is plotted versus $z$ for two different types of observations, SNe and GRB; and these observations are compared to the three different models. ($\mu$ is a normalized measure of the distance to an observed object.) We note that all three models provide a reasonably good fit to some of the data in Fig.~\ref{PlotsSNeGRB}, but there are also differences between the three curves that are clearly visible, and if the plot were scaled in (zoomed in) for low redshift, we would see more differences.    For a model with more matter (more gravity), the matter density would use gravity to try to pull the universe together, subsequently slowing the expansion rate, so the objects in the universe would be closer together (bottom curve).  Conversely, a universe with too much dark energy would expand the universe too quickly, and objects would be at greater distances than what is observed (top curve).  

\begin{figure}[h!]

\begin{center}

\includegraphics[scale=0.28]{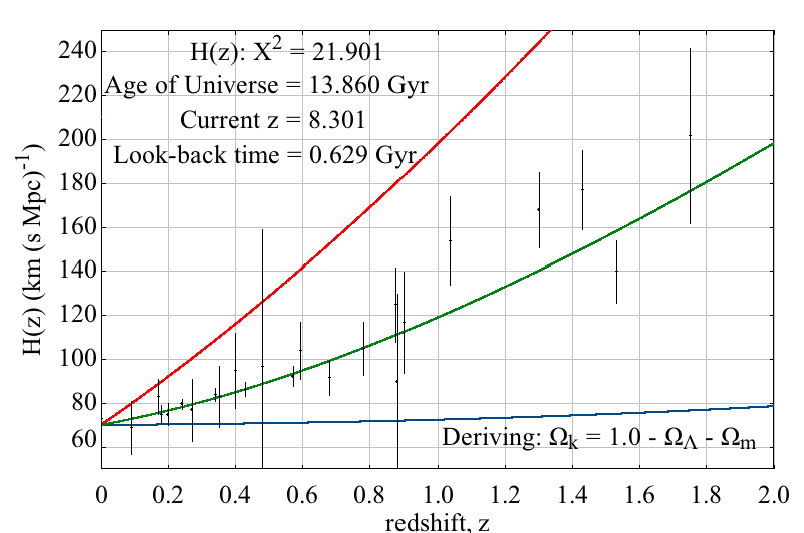}
\caption{\label{PlotsHz}  (color online)
Hubble Parameter, $H(z)$ versus redshift. \texttt{COSMOEJS} output showing three different models (curves) compared to a data set of the expansion rates $H(z)$ of galaxies at different redshift.  The three models differ \emph{only} in their fractional matter and dark energy densities, $\{\Omega_m,\,\Omega_\Lambda\}=\{0.01,\,0.99\},\,\{0.27,\,0.73\},\,\{1.0,\,0.0\}$ correspond to bottom (blue), middle (green), and top (red), respectively.  (Note: Labels are draggable in \texttt{COSMOEJS}.) }

\end{center}
\end{figure}

In Fig.~\ref{PlotsHz}, the expansion rates, $H(z)$, of different galaxies are plotted as a function of their redshift, $z$. These data are compared with the same three models that are used for Fig.~\ref{PlotsSNeGRB}, and again, 
all three the models have been defined to have the same expansion rate today, i.e., $H(z) \equiv H_0$ for $z=0$.  From these data it is clear that the middle curve ($\Omega_m = 0.27$, $\Omega_\Lambda = 0.73$) gives a better fit than the other two models.  This plot correctly displays that a model universe comprised of mostly dark energy (bottom curve) would have a slightly increasing expansion rate as a function of redshift for the range of redshift seen in Fig. \ref{PlotsHz}, whereas a universe with mostly matter would have an expansion rate that drastically decreases with increasing time (decreasing $z$). 

\begin{figure}[h!]

\begin{center}

\includegraphics[scale=0.28]{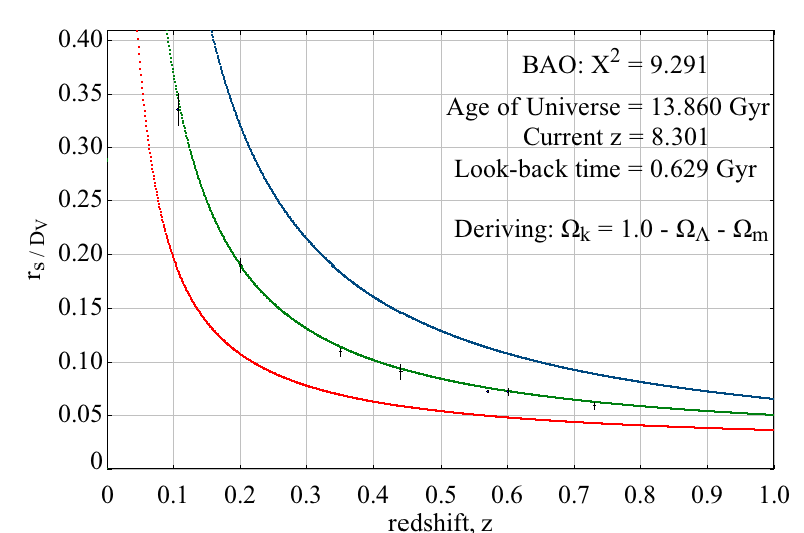}

\caption{\label{PlotsBAO}  (color online)
Baryon Acoustic Oscillations versus redshift. \texttt{COSMOEJS} output showing three different models (curves) compared to data sets of the BAO ratio, the size of the sound horizon, $r_s$, to its effective distance, $D_v$ in the galaxies.  The three models differ \emph{only} in their fractional matter and dark energy densities, $\{\Omega_m,\,\Omega_\Lambda\}=\{0.01,\,0.99\},\,\{0.27,\,0.73\},\,\{1.0,\,0.0\}$ correspond to top (blue), middle (green), and bottom (red), respectively.  (Note: Labels are draggable in \texttt{COSMOEJS}.)}

\end{center}
\end{figure}

In Figure \ref{PlotsBAO}, the three models show a \textit{clear} difference when compared to the BAO ratio.  Physically, the BAO ratio reflects the size of the sound horizon, $r_s$, for the early universe baryon decoupling (see subsection \ref{sec:expobservations} of the Appendix) to its effective distance, $D_v$ in the galaxies today.   In Figs. \ref{PlotsSNeGRB} and \ref{PlotsHz}, the two extreme models (upper and lower curves) at least fit some of the data, but they do not come close to the extremely precise (small error bars) BAO ratio data.  With the addition of this third observation, it is obvious that only the middle curve fits well to all of the complementary data sets.  (By ``complementary," we mean that the cosmological parameters must be consistent with different observations that constrain different theoretical observables.) 

 The user can easily vary additional parameters, subsequently adding to the complexity of the models, by simply adjusting the sliders for the different parameter values.  For example, by varying the value of the current expansion rate, $H_0$, while looking at the same compositions, the changing of the expansion rate will uniformly scale all of the theoretical data points. In addition, the user can choose different spatial curvatures for the universe, $\Omega_k$, and different types of dark energy $\{w_0, w_a\}$ models.


\subsection{Cosmological interpretation}\label{sec:interpretation}
For the cosmological interpretation of these fits, we proceed to plotting the evolution of the expansion of the universe versus to time, $t$ (in Gyrs $=10^9$ years), rather than redshift, $z$. Redshift is a model-independent measurement, which makes it an ideal quantity for the fitting that was done in Sec.~\ref{sec:fitting}.  However, redshift has a non-linear, model-dependent relationship with time, which makes it very difficult to physically interpret data that are plotted versus $z$.\footnote{The relationship between time and redshift (through $a(t)$) can only be calculated once a specific model is chosen.}  For this reason, we now plot the expansion of the universe versus time for each of the three models that were studied in Sec. ~\ref{sec:fitting}.  Specifically, we plot the dimensionless ratio $a(t)/a(t_{today})$, where $a(t)$ represents the expansion function or radius (size) of the universe.  It is also referred to as the ``scale factor," and mathematically, it is defined as $a=1/(1+z)$.  \footnote{Actually, we evolve the scale factor, $a$, and calculate the time, $t$ (in Gyrs $=10^9$ years), but switch the $x$ and $y$ axis, since this is the more common convention.}

 \begin{figure}[h!]

\begin{center}

\includegraphics[scale=0.28]{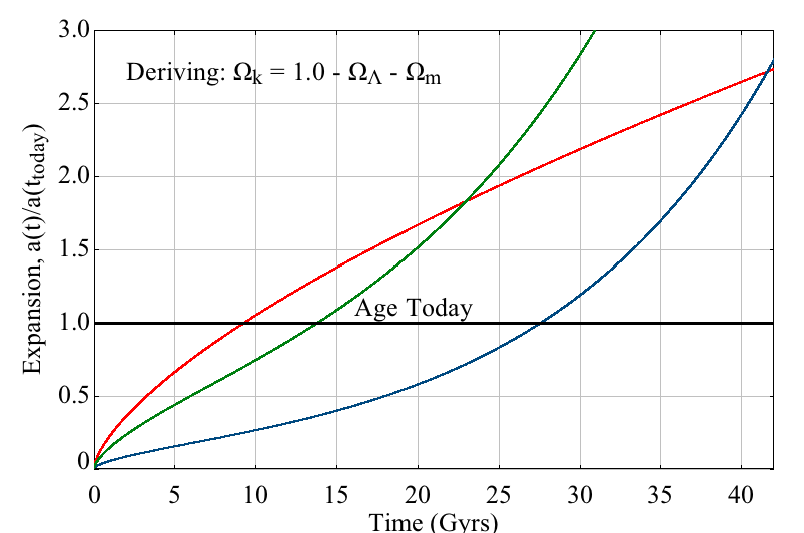}

\caption{\label{PlotsExp}  (color online)
Expansion, $a(t)/a(t_{today})$ versus time (Gyrs $=10^9$ years). \texttt{COSMOEJS} output showing three different models (curves) of the expansion of the universe, $a(t)/a(t_{today})$ (scale factor), versus time.  The three models differ \emph{only} in their fractional matter and dark energy densities, $\{\Omega_m,\,\Omega_\Lambda\}=\{0.01,\,0.99\},\,\{0.27,\,0.73\},\,\{1.0,\,0.0\}$ correspond to bottom (blue), middle (green), and top (red), respectively below the solid ``Age Today" (black) line.  (Note: Labels are draggable in \texttt{COSMOEJS}.)}

\end{center}
\end{figure}

In Figure \ref{PlotsExp}, we provide plots of all three models studied in Sec.~\ref{sec:fitting}.  These plots can be interpreted to give physical insight to the expansion at a particular time.  We know from the fitting in Sec.~\ref{sec:fitting} that the model with \textit{all} matter density did not match the data, and if we look at its expansion, we can see the slope decreases monotonically with increasing time, so the rate of the universe's expansion is decreasing, which would be caused by the gravitational pull of the matter dominating the universe. This slowed expansion yields an age of only $\sim9$ Gyrs.  For the other extreme case studied in Sec.~\ref{sec:fitting}, we see a very early inflection point where the expansion rate went from decreasing to accelerating, yielding a much older universe of $\sim28$ Gyrs.  This acceleration is caused by the domination of the dark energy as the universe expands.  Finally, for the model which fit all of the observations in Sec.~\ref{sec:fitting}, we see an inflection point at $\sim9$ Gyrs, which is caused by the domination of the dark energy as the universe expands and the gravitational pull from matter is weakened.  This model calculates a current age of the universe to be $\sim14$ Gyrs.

 According to the combinations of observations in the current literature \cite{CurrentData, AndersonBAOBOSS, HzClocks, WMAP2010}, the best-fit values for the parameters in \texttt{COSMOEJS} are $\{H_0\sim70.0$ km/(s Mpc), $\Omega_b\sim0.04$, $\Omega_c\sim0.23$, $\Omega_m\sim0.27$, $\Omega_\Lambda\sim0.73$, $\Omega_k\sim0.0$, $w_0\sim-1.0$, $w_a\sim0.0\}$, representing the so-called $\Lambda$ Cold Dark Matter ($\Lambda$CDM) model.  With \texttt{COSMOEJS}, it is possible to find these values by systematically trying different sets of parameters with combinations of the data sets.  Then, a physical interpretation of the model's fit throughout its evolution can be made to compare with the cosmological observations \footnote{It is important to compare the model both numerically and visually inspecting the plots, but also make sure the model is physical (e.g. $\Omega_m= 0$ is unphysical, due to no matter).}.
 

\section{Conclusions and Future Work}\label{sec:summary}

\texttt{COSMOEJS} is a powerful new tool for cosmology education, and it is also precise enough to perform research grade calculations for testing most cosmological dark energy models. They also allow the user to select inputs for parameters that are perhaps not scientifically accepted. This allows the user to discover how parameters influence the shape of the curve for a particular theoretical model, thereby understanding the physical interpretation of a model's fit to the data. Variations of the programs have been used for science outreach and for classroom illustration.  

Future versions of the programs will involve a minimization method for the fitting of the cosmological models to the data sets to provide best-fit cosmological parameters. However, this will involve a different fitting method for each survey.  We decided not to provide minimization in this version because this would distract from the pedagogical value of the program.  Namely, when trying to find a best-fit model with a minimization routine, the user is not required to understand the physical interpretation of one fit over another.  The fit is obtained by statistically comparing one model fit to another.  Also, if the statistical fit is biased in some way, as explained in the examples above for $\chi^2$ fits, then the best fit could have unphysical parameter constraints.  In the current version of the simulation, we are more concerned with understanding the physical interpretation of fitting particular cosmological models to data sets.  

\texttt{COSMOEJS} allows non-specialists to manipulate cosmological models via their parameters and learn how to fit the model to experimental data sets. This manual process of changing parameter values also allows the user to see what parameter values do not fit the data.  The programs are useful for not only learning about cosmology but also data fitting itself, both visually and numerically.  The programs will continue to receive updates and modifications for new, more precise data sets as these become publicly available.  Using the $\Lambda$CDM model with this version of \texttt{COSMOEJS}, we find excellent fits to all the data sets with $\{H_0=70.0$ km/(s Mpc), $\Omega_b=0.045$, $\Omega_c=0.225$, $\Omega_m=0.27$, $\Omega_\Lambda=0.73$, $\Omega_k=0.0$, $w_0=-1.0$, $w_a=0.0\}$.
\begin{appendix}
\section{Cosmological Background}\label{sec:background}
In this appendix, we include a brief tutorial into cosmology. In subsection \ref{sec:mathematics} of the Appendix, we discuss the mathematics of the theoretical cosmology in \texttt{COSMOEJS}, followed by a description of the observations in subsection \ref{sec:expobservations} of the Appendix. For a more detailed description of cosmology, see the supplementary information in Ref.~\onlinecite{Observations}.

The cosmic acceleration of the universe can be explained by a cosmological constant, or some other form of repulsive dark energy, \textit{i.e.} a negative pressure and a negative equation of state, or by an extension or modification to gravity at cosmological scales of distances \cite{Reviews}.  In the context of general relativity (GR), to account for this dark energy effect, the addition of a $\Lambda$ term (cosmological constant) to Einstein's Field Equations (EFE) can be used to derive equations of motion with a cosmological constant of the desired value consistent with the dynamics of Friedmann-Lemaitre-Robertson-Walker (FLRW) universe \footnote{A detailed derivation of these models and an analysis for some special cases was recently made available \cite{PhysicsToday1, PhysicsToday2, SonegoTalamini}.}.  We provide a means of testing this commonly accepted model of the Universe and others with observations of SNe Type Ia, gamma-ray bursts (GRB), baryon acoustic oscillations (BAO), the distance to the last scattering surface of the cosmic microwave background (CMB) radiation, and measurements of the Hubble expansion rate parameter, $H(z)$, thereby deriving the parameters for the standard model in cosmology.

\subsection{Mathematics}\label{sec:mathematics}
In this section, we define the mathematics behind the theoretical models involved in \texttt{COSMOEJS}.
Specifically, the programs assume a big bang physical universe, a mathematical model according to general relativity (GR), and a uniformly distributed spacetime in all directions \footnote{In cosmology, the Friedmann-Lemaitre-Robertson-Walker (FLRW) spacetime metric \cite{Rindler} describes a homogenous and isotropic spacetime.}.   From these assumptions, the programs numerically integrate an equation of motion for the dynamical evolution of the expansion rate of the universe \footnote{When using the FLRW metric, the equation of motion is referred to as the Friedmann equation.}.  This equation of motion can be expressed in terms of the Hubble expansion rate, $H(z)$, as a function of redshift, $z$. [The theoretical details of the integration of $H(z)$ for a particular observation are described in Ref.~\onlinecite{Observations}, and the numerical implementation is shown in Fig. 1 of the appendix in Ref.~\onlinecite{sec:ejs}.]  Explicitly, we use
\begin{widetext}
\be 
H(z)=H_0\sqrt{\Omega_m(1+z)^3+\Omega_\Lambda\Big[(1+z)^{3(1+w_0+w_a)}\exp\Big(\frac{-3 w_a z}{1+z}\Big)\Big] + \Omega_k (1+z)^2}.
\label{eq:FriedmannEquationH}	
\ee
\end{widetext}
This equation contains all of the parameters that can be varied in \texttt{COSMOEJS}.  (All parameters are dimensionless except $H_0$).  Briefly, these parameters and their currently accepted values are:
\begin{itemize}
\item $H_0 \approx 70.0$ km/(s Mpc) : the Hubble Constant;
\item $\Omega_m \approx 0.27$ : the fractional matter density (subject to the constraint: $\Omega_m = \Omega_b + \Omega_c$);
\item $\Omega_b \approx 0.04$ : the fractional baryon density;
\item $\Omega_c \approx 0.23$ : the fractional cold dark matter density;
\item $\Omega_\Lambda \approx 0.73$ : the fractional dark energy density;
\item $\Omega_k \approx 0.0$ : the fractional curvature density;
\item $\Omega_0 \approx 1.0$ : the sum total energy density (subject to the constraint: $\Omega_0 = \Omega_m + \Omega_\Lambda + \Omega_k$);
\item $w_0 \approx -1$ : the equation of state of dark energy ;
\item $w_a \approx 0.0$ : the derivative of $w_0$; 

\end{itemize}
As the values of these parameters change, Eq. (\ref{eq:FriedmannEquationH}) describes different types of evolutions for the universe.  The details of these parameters are further explained in the next paragraph.

\begin{figure}[h!]
\begin{center}
\includegraphics[width= 6.5cm]{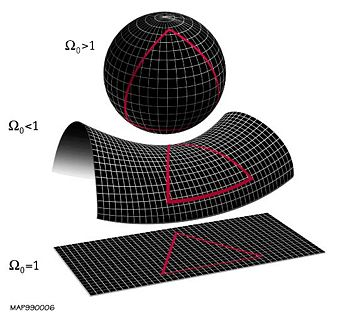}
\caption{\label{Geoms}  (color online)
 Universal geometries (assuming isotropy and homogeneity) for illustrating various inherent spacetime geometries given by $\Omega_0$ ($\Omega _k$) \cite{NASAWMAPTeam}.  TOP: Closed, i.e. sum of angles in triangle is greater than $180^o$; MIDDLE: Open, i.e. sum of angles in triangle is less than $180^o$; BOTTOM: Flat, i.e. sum of angles in triangle is exactly $180^o$.  Image used with permission from NASA WMAP Science Team (2012), $<$\url{http://map.gsfc.nasa.gov/media/}$>$.}
\end{center}
\end{figure}

The Hubble Constant parameter, $H_0$, represents the current value of the expansion rate for the universe. The $k$ in $\Omega_k$ appearing in Eq. (\ref{eq:FriedmannEquationH}) represents the three types of curvature for the spacetime of the universe as open, flat, or closed ($k=-1,\,0,\,1$, respectively).  See Fig. \ref{Geoms}. This is an inherent curvature of the empty spacetime itself, devoid of any matter or energy.  However, as can be seen, the model does not use $k$ directly, but rather the fractional curvature density parameter, $\Omega_k$.  The total energy density of the universe is split up into fractional pieces to represent its different compositional quantities for matter, dark energy and curvature.  The total energy density, $\Omega_0$, as measured today ($z=0$), accounts for the sum total of all of the matter and energy in the universe.  A critically dense universal model (typically accepted in cosmology\footnote{Recent data shows $\Omega_0\sim1$ \cite{CurrentData, AndersonBAOBOSS, HzClocks, WMAP2010}, although a more precise survey \cite{PLANCK} may uncover a slight offset which can be modified in the source code, see supplementary information in Ref.~\onlinecite{sec:ejs}}), indicating that all matter and energy are accounted for, can be described by the relation

\be 
\Omega_0 = \Omega _m + \Omega _\Lambda + \Omega _k = 1.
\label{eq:CriticalDensity}
\ee

Note the curvature parameter, $k$, has a negative sign ($-$) originating from the GR spacetime equations of motion \cite{Rindler}.  However, the fractional curvature parameter, $\Omega_k$, has the relationship \cite{BergstromGoobar}
\be 
\Omega _k \propto -k,
\label{eq:CurvatureParameter}
\ee
such that

\begin{displaymath}
   k = \left\{
     \begin{array}{lr|r}
       -1 & : \Omega _k > 0 &\,\textnormal{Open\, (negative\, curvature)}\\
        0 & : \Omega _k = 0 &\,\textnormal{Flat\, (zero curvature)}\\
        1 & : \Omega _k < 0 &\,\textnormal{Closed \,(positive\, curvature)}.
       
     \end{array}
   \right.
\end{displaymath}

The fractional matter density, $\Omega_m$, represents the total matter density in the universe.  It can be separated into its constituents, $\Omega_m = \Omega_b + \Omega_c$, into fractional baryon density, $\Omega_b$, and fractional cold dark matter density, $\Omega_c$, when the observation can constrain the distinction (only BAO and CMB observations \cite{DataTable}).  Note, for all values of $\Omega_m$ that were used in Sec.~\ref{sec:modeling}, we kept the constituents the same proportionate percentages of $\Omega_m$ as the currently accepted values listed above.
The $\Lambda$ represents the cosmological constant, the simplest model of dark energy.
For changing the model of dark energy, the model uses an equation of state parameterized as $w(z) = w_0+w_a[z/(1+z)]$ \cite{Polarski2000}.  The equation of state for dark energy includes, $w_0$, which is the value measured today ($z=0$), and its derivative, $w_a$, that allows dark energy to evolve in time, $z$, as the universe evolves.   For the special condition when $w_a=0.0$, the equation of state is constant, i.e. the density of dark energy does not change with time.  Additionally, when $w_0=-1$, this corresponds to the cosmological constant model.

\subsection{Experimental Observations}\label{sec:expobservations}
Within the \texttt{COSMOEJS} package, we include 18 different experimental datasets
for five different types of measurements: Type Ia supernovae (SNe),
gamma-ray bursts (GRB), measurements of the Hubble expansion rate
parameter, $H(z)$, baryon acoustic oscillations (BAO), and the cosmic
microwave background (CMB) radiation.  In cosmology, the evolution of the universe is modeled on scales too large to measure the evolution of a single galaxy or galaxy cluster from its formation to the present.  Instead, cosmology combines observations of different galaxies---and the phenomena contained therein---at different times and distances in their evolution to piece together the dynamics of the universe. These different observations and surveys provide independent and complementary measurements of the expansion history of the universe and its composition.  

According to the big bang theory, the universe began from an initial state of extremely high temperature, density and energy.  When the universe had expanded and cooled enough for the photons to decouple from the primordial soup of energy, they no longer scattered and were free to propagate throughout the universe, allowing for their detection.  This surface is as far back as scientists can currently make measurements because all the measurements involve some frequency of light.  These high energy photons have been stretched with the expansion of the universe into microwaves.  So, now the temperature of the universe has cooled to $\approx 3$ K.  Working backwards, this gives a temperature of $\approx 380,000$ K for the last scattering surface.  We use the three fitting parameters for amplitude, and locations of the acoustic peaks of the CMB temperature power spectrum \cite{WMAP2010}: 1) the acoustic scale, $l_a(z_*)$, 2)  the shift parameter, $R(z_*)$, 3) the redshift of the surface of last scattering (SLS) of the CMB, $z_*$. (For an accessible introduction to the CMB power spectrum and how temperature oscillations become acoustic peaks, see Ref.~\onlinecite{WayneHu}.) Physically the acoustic scale and shift parameter are the size, shape and position of the acoustic peaks of the CMB power spectrum for different values of the SLS redshift. The size, shape and positions of these acoustic peaks are very powerful in determining the values of the cosmological parameters, due to the complexity of the many peaks in the CMB power spectrum and their ratios to each other.

 The SNe are standard ``candles" (similar luminosity) used to form a redshift-distance relation, $\mu(z)$ (distance modulus),  to measure the rate of the expansion of the universe.   According to these measurements, galaxies at large distances, in which the SNe reside, are receding less rapidly than Hubble's law would predict.   This translates to a slower expansion rate in the past, and that the nearby, later time SNe are expanding faster than the more distant, older SNe.  Therefore, we are observing an accelerating universe.  GRBs are added to fill the large void of redshift between the high-$z$ SNe and the redshift of the CMB's last scattering surface, $z\approx 1089$.  SNe are subject to extinction from the dust of the interstellar medium, however, GRBs are much brighter and, due to the high energy of gamma-ray photons, are rarely affected by the dust. The measurement of the GRB extends the redshift-distance relationship to higher redshift, although, there is a redshift range of overlapping measurements for comparison and consistency.

The measurements of the Hubble Parameter, $H(z)$ are an independent measurement of the expansion history of the universe.  All of the other cosmological observations given in this article require an integration, but $H(z)$ is an exact evaluation and comparison of Eq. (\ref{eq:FriedmannEquationH}) to the experimental data.  In fact, $H(z)$ is actually a direct measurement of the differential age of the universe, $\Delta z/\Delta t$, in other words measuring how the age of the universe changes as the redshift changes using the age differences of old elliptical galaxies that are passively evolving \cite{HzClocks}.  They are used as standard ``clocks" to directly probe the Hubble parameter. 

We compare the ratio of the sound horizon at the drag epoch, $r_s(z_d)$, or when the baryons decoupled from the primordial universe to its effective distance, $D_v(z)$ in the galaxy redshift surveys.  This decoupling occurs at a slightly later time and lower redshift than the photon decoupling because the baryons get ``stuck" in gravitational potential wells.  The correlations in the galaxy redshift surveys consistently have a ``bump" corresponding to the standard ``ruler" measurement of the BAO.  This measures the expansion of the primordial sound horizon in the galaxy redshift surveys.   As an example of this physical ratio, the sound horizon at decoupling in the range of $r_s=153.19$ Mpc, and effective distance $D_v(z=0.57)=2026.49$ Mpc, for a ratio of $r_s/D_v=0.076$, \cite{AndersonBAOBOSS}. Physically, this reflects the size and shape of the acoustic peak, and how it has evolved with the expansion of the universe. The BAO is specially suited for constraining $\{\Omega_b,\,\Omega_c\}$ with galaxy clusters because of the sensitivity of the sound horizon redshift to these parameters.

 Equation (\ref{eq:FriedmannEquationH}) can be used to compare several dark energy cosmological models to these observations by allowing different values for the parameters, $\{\Omega_m\, (=\Omega_b+\Omega_c),\,\Omega_\Lambda,\,\Omega_k, H_0,\, w_0,\,$ and $w_a\}$.  While a model may fit one observation, cosmology involves the entire evolution of the universe, so it is important to use all the cosmological observations, covering several redshift and cosmological epochs, to find a best-fit model.

\end{appendix}
\acknowledgments
{The authors wish to thank W. Christian and F. Esquembre for adding new features to EJS for these programs upon request. We also thank D. Jokisch for useful comments. JM would especially like to thank M. Ishak, W. Rindler and The University of Texas at Dallas Cosmology, Relativity and Astrophysics group for his training in computational and theoretical cosmology. Part of this work was completed under the SC Space Grants Palmetto Academy Program, Award  NNG05G168G. }

\end{document}